\documentclass[letterpaper, 10 pt, conference]{ieeeconf}  % Comment this line out if you need a4paper

\IEEEoverridecommandlockouts                              % This command is only needed if 
\usepackage{graphicx} % for pdf, bitmapped graphics files
\usepackage{amsmath} % assumes amsmath package installed
\usepackage{amssymb}  % assumes amsmath package installed
\usepackage{cite}
\usepackage{tabularx} 
\title{\LARGE \bf
Traffic State Prediction and Control Using a Koopman Operator Framework with Polynomial Observables}

\author{Shakib Mustavee (Member IEEE)$^{1}$, Arvind Singh$^{2}$, and Shaurya Agarwal (Senior Member IEEE)$^{3}$% <-this % stops a space
\thanks{*A. Singh acknowledges partial support from the U.S. National Science Foundation under Grant EAR-2342936.}% <-this % stops a space
\thanks{$^{1}$Shakib Mustavee is a Post-doc with the Department of Civil, Environmental \& Construction Engineering, University of Central Florida, Orlando, Florida.
        {\tt\small shakib.mustavee@ucf.edu}}%
\thanks{$^{2}$ Arvind Singh is an Associate Professor with the Department of Civil, Environmental \& Construction Engineering, University of Central Florida, Orlando, Florida.
        {\tt\small arvind.singh@ucf.edu}}%        
\thanks{$^{3}$ Shaurya Agarwal is an Associate Professor with the Department of Civil, Environmental \& Construction Engineering, University of Central Florida, Orlando, Florida.
        {\tt\small shaurya.agarwal@ucf.edu}}%
}

\begin{document}

\maketitle
\thispagestyle{empty}
\pagestyle{empty}

\begin{abstract}
In this paper, we develop a Koopman operator framework for modeling and control of traffic flow governed by the Lighthill--Whitham--Richards (LWR) conservation law. We show that the nonlinear dynamics induced by the Greenshields flow--density relation naturally motivate a dictionary of polynomial observables, while the controlled dynamics become bilinear in the lifted Koopman coordinates. A reduced-order Koopman model is identified from spatio-temporal traffic-density snapshots and used as the prediction model in a model predictive control (MPC) framework for ramp metering. The resulting Koopman-MPC is compared with two predictive control baselines that use, respectively, a fixed linearized LWR model and the nonlinear finite-volume LWR model. In a one-hour freeway experiment, the Koopman-based controller maintains the traffic density below the critical Greenshields value and closely reproduces the behavior of the nonlinear PDE-based MPC, with a maximum-density difference below $1\%$. The Koopman predictor also reduces the mean online model-evaluation cost by approximately one order of magnitude relative to the nonlinear PDE predictor. In contrast, the linearized LWR predictor is computationally cheaper but exhibits substantially larger prediction errors and yields a more conservative metering policy. These results demonstrate that the analytical structure of the nonlinear traffic conservation law can be exploited to construct an interpretable data-driven surrogate for efficient traffic prediction and control.
\end{abstract}

\section{Introduction}
\label{sec:introduction}

At the macroscopic scale, traffic flow is described in terms of aggregate quantities such as density, flow, and velocity, providing a continuum representation of traffic dynamics. Among macroscopic traffic models, the Lighthill--Whitham--Richards (LWR) model is one of the most widely used\cite{LighthillWhitham1955,Richards1956}. It describes the evolution of vehicle density through a nonlinear scalar conservation law. The LWR model can capture important traffic phenomena such as shock-wave formation and propagation, and queue evolution. Therefore, the model has become a basic building block for traffic-state prediction and control. However, the nonlinear and hyperbolic nature of the governing partial differential equation (PDE) makes repeated prediction computationally demanding in feedback-control applications. In model predictive control (MPC), for example, candidate control actions must be evaluated by repeatedly propagating the traffic model over a prediction horizon. This motivates reduced and data-driven representations that retain the essential nonlinear traffic dynamics while providing more efficient prediction models for online control.

Koopman operator theory provides one such representation by describing a
nonlinear dynamical system through the linear evolution of functions of the
state, referred to as observables. Although the Koopman operator is generally
infinite-dimensional, finite-dimensional approximations can be constructed
from data. Extended dynamic mode decomposition (EDMD) provides a systematic
framework for approximating the Koopman operator using a prescribed dictionary
of observables \cite{Williams2015EDMD}, while generator EDMD (gEDMD) extends
this viewpoint to approximation of the infinitesimal Koopman generator
\cite{Klus2020gEDMD}. These representations are particularly attractive for
control because nonlinear dynamics can be lifted into coordinates in which
prediction is performed using linear or bilinear evolution operators.
Koopman-based predictors have consequently been incorporated into MPC for
general nonlinear systems \cite{KordaMezic2018}, including nonlinear PDEs.

Koopman- and DMD-based methods have been applied to a range of traffic problems, including highway traffic flow analysis, signalized traffic prediction and control, modeling queue evolution at intersections, and capturing microscopic and multi-lane traffic dynamics \cite{avila2020data, Ling2022TrafficKoopman,DasMustaveeAgarwalHasan2023,
ShababMustaveeAgarwal2024,MustaveeAgarwal2025}. Koopman representations have also gained increasing attention in model-based control of traffic flow because the lifted dynamics can serve as computationally efficient predictors within optimization-based control frameworks. For example, Gu \emph{et al.} integrated a deep Koopman model with MPC for freeway ramp metering \cite{GuZhouWu2023}, and more recent work extended Koopman-based latent dynamics to model-based planning for ramp-metering control \cite{ZhouGuLimYuan2026}. Xiang \emph{et al.} incorporated LWR-based density evolution into a Koopman framework that couples macroscopic traffic flow with microscopic vehicle dynamics for traffic prediction \cite{XiangZhaoYu2026}. In addition, Koopman theory was combined with graph-based learning to model and predict nonlinear spatio-temporal highway traffic dynamics \cite{WangNgoduy2024}.

\noindent \textit{Research gap:} Despite this progress, the analytical
connection between the classical LWR conservation law and the construction of
Koopman observables has received considerably less attention. Existing traffic-oriented Koopman studies have largely relied on DMD, learned latent coordinates, or neural-network-based lifting. The observable space has generally not been derived directly from the analytical structure of the governing traffic PDE.
Operator-theoretic representations of nonlinear PDEs have been studied in
other settings, including Burgers' equation \cite{PageKerswell2018}. However,
to the best of our knowledge, the polynomial observable structure induced by
the LWR model with Greenshields flux has not been explicitly derived and
exploited for traffic control.

\noindent \textit{Novelty:} In this work, we identify suitable Koopman observables using the mathematical structure of the LWR model and the Greenshields flow--density relation. The quadratic flux naturally leads to polynomial functions of traffic density and provides a direct basis for the lifting functions. We derive their evolution equations, quantify the finite-order truncation error, and extend the formulation to ramp-metering control, yielding a bilinear reduced-order model identified from traffic-density snapshots. The model is used within a receding-horizon controller and compared with MPC based on a fixed linearization and the full nonlinear finite-volume LWR model under identical control settings.

\noindent \textit{Contributions:} The main contributions of this paper are summarized as follows:
\begin{itemize}
    \item We derive the evolution equations for polynomial functions of traffic density directly from the LWR model by exploiting the Greenshields flow--density relation, thereby providing an analytical basis for selecting the Koopman observables.

    \item We characterize the truncation residual introduced by a finite
    polynomial basis and show that it vanishes asymptotically for smooth traffic states with density strictly below the jam density.
    
    \item We extend the formulation to ramp-metering control and identify a
    reduced bilinear Koopman predictor from traffic-density snapshots. The
    resulting MPC is compared with controllers based on a fixed linearized
    LWR model and the full nonlinear finite-volume LWR model under identical
    control settings.
\end{itemize}

The remainder of the paper is organized as follows. Section~\ref{sec:koopman_lwr} develops the Koopman representation of the LWR model with the Greenshields flow--density relation, discusses the finite-order approximation, and derives the controlled bilinear model used for receding-horizon control. Section~\ref{sec:case_study} presents the ramp-metering case study, including validation of the learned Koopman predictor and comparison with MPC formulations based on a fixed linearized LWR model and the nonlinear finite-volume LWR model. Finally, Section~\ref{sec:conclusion} summarizes the main findings and discusses directions for future work.

\section{Mathematical Background and Derivations}
\label{sec:koopman_lwr}

\subsection{Koopman Observables for the LWR Model}
\label{subsec:poly_koopman}
The LWR model describes the evolution of traffic density
$\rho(x,t)$ through the scalar conservation law
\begin{equation}
    \frac{\partial \rho}{\partial t}
    +
    \frac{\partial Q(\rho)}{\partial x}
    =0,
    \label{eq:lwr_general}
\end{equation}
where $Q(\rho)$ denotes the equilibrium flow--density relation. Under the Greenshields velocity model,
\[
    V(\rho)
    =
    v_f
    \left(
        1-\frac{\rho}{\rho_j}
    \right),
\]
the corresponding traffic flux is $Q(\rho)=v_f\rho(1-\rho/\rho_j)$, where $v_f$ and $\rho_j$ denote the free-flow speed and jam density, respectively. Introducing the normalized traffic density $r=\rho/\rho_j$, the normalized LWR equation becomes
\begin{equation}
    r_t
    +
    v_f\partial_x
    \left[
        r(1-r)
    \right]
    =0,
    \label{eq:lwr_normalized}
\end{equation}
or, equivalently,
$r_t=-v_f r_x+v_f\partial_x r^2$. 

For sufficiently smooth solutions of \eqref{eq:lwr_normalized}, we consider
the pointwise monomial observables
\[
    \psi_m(r)=r^m,
    \qquad
    m=1,2,\ldots,
\]
which together define the polynomial Koopman dictionary. The temporal
evolution of each observable is obtained along trajectories of
\eqref{eq:lwr_normalized}. Applying the chain rule gives
\begin{align*}
    \partial_t r^m
    &=
    m r^{m-1}r_t
    \\
    &=
    -m v_f r^{m-1}r_x
    +
    2m v_f r^m r_x .
\end{align*}
Using
$\partial_x r^m=m r^{m-1}r_x$ and
$\partial_x r^{m+1}=(m+1)r^m r_x$, we obtain
\begin{equation}
    \partial_t r^m
    =
    -v_f\partial_x r^m
    +
    \frac{2m}{m+1}
    v_f\partial_x r^{m+1}.
    \label{eq:polynomial_hierarchy}
\end{equation}

Equation~\eqref{eq:polynomial_hierarchy} shows that the quadratic
Greenshields flux naturally induces a polynomial dictionary of observables.
In particular, the evolution of $r^m$ depends on the spatial derivatives of
$r^m$ and the next-degree observable $r^{m+1}$. The first few relations are
\begin{align*}
    r_t
    &=
    -v_f r_x
    +
    v_f\partial_x r^2,
    \\
    \partial_t r^2
    &=
    -v_f\partial_x r^2
    +
    \frac{4}{3}v_f\partial_x r^3,
    \\
    \partial_t r^3
    &=
    -v_f\partial_x r^3
    +
    \frac{3}{2}v_f\partial_x r^4 .
\end{align*}

Defining the infinite observable vector \(
    \boldsymbol{\Psi}
    =
    \begin{bmatrix}
        r &
        r^2 &
        r^3 &
        \cdots
    \end{bmatrix}^{\top},
\)
these relations may be written compactly as
\begin{equation}
    \partial_t\boldsymbol{\Psi}
    =
    C\,\partial_x\boldsymbol{\Psi},
    \label{eq:continuous_operator_form}
\end{equation}
where
\[
    C
    =
    v_f
    \begin{bmatrix}
        -1 & 1 & 0 & 0 & \cdots \\
        0 & -1 & \frac{4}{3} & 0 & \cdots \\
        0 & 0 & -1 & \frac{3}{2} & \cdots \\
        \vdots & \vdots & \vdots & \ddots & \ddots
    \end{bmatrix}.
\]
Thus, on this family of observables, the action of the Koopman generator may
be written formally as
\[
    \mathcal{L}=C\,\partial_x .
\]
The matrix $C$ describes the coupling among the polynomial observables,
whereas $\partial_x$ accounts for the spatial transport in the conservation law. For a polynomial dictionary truncated at degree $p$,
\(
    \boldsymbol{\Psi}_p
    =
    \begin{bmatrix}
        r &
        r^2 &
        \cdots &
        r^p
    \end{bmatrix}^{\top},
\)
the first $p-1$ observable equations remain within the retained space. The
highest-order equation, however, depends on the unresolved observable
$r^{p+1}$, giving the truncation residual
\begin{equation}
    \boldsymbol{\mathcal R}_p
    =
    \frac{2p}{p+1}
    v_f
    \partial_x r^{p+1}
    \mathbf e_p
    =
    2pv_f r^p r_x\,\mathbf e_p ,
    \label{eq:closure_residual}
\end{equation}
where $\mathbf e_p$ is the $p$-th canonical basis vector. For smooth traffic
states satisfying
\[
    0\leq r(x,t)\leq\bar r<1,
    \qquad
    \|r_x\|_{\infty}\leq M,
\]
the residual satisfies
\begin{equation}
    \|\boldsymbol{\mathcal R}_p\|_{\infty}
    \leq
    2v_f M p\bar r^p
    \rightarrow0
    \qquad
    \text{as } p\rightarrow\infty .
    \label{eq:closure_bound}
\end{equation}
Thus, for traffic states without shock discontinuities, with bounded spatial gradients and density strictly below the jam density, the unresolved contribution vanishes asymptotically as the polynomial order increases. This bound characterizes the truncation of the continuous polynomial hierarchy and does not imply a monotonic decrease in the prediction error of the learned finite-dimensional model.

To connect the continuous representation with a finite-dimensional numerical model, the spatial differential operator must also be discretized. For example, if $D_x$ denotes a consistent discrete differentiation operator, then the grid-point values of $\partial_x r^m$ may be approximated by $D_x\mathbf r^{\circ m}$, where $\mathbf r=[r_1,\ldots,r_N]^{\top}$ denotes the discretized density field at $N$ spatial locations and $\mathbf r^{\circ m}=[r_1^m,\ldots,r_N^m]^{\top}$ denotes its elementwise $m$th power. For sufficiently smooth solutions, consistency implies that this approximation approaches the corresponding spatial derivative as the grid is refined \cite{LeVeque2002}. Under such a linear spatial discretization, the truncated operator $C\partial_x$ would therefore be represented by a finite block operator constructed from the polynomial coupling and the discrete spatial derivative. In this study, the nonlinear LWR equation is propagated using a finite-volume numerical flux without explicitly constructing the spatial operator from a differentiation matrix. The polynomial observables are evaluated from the resulting traffic-density fields, and the reduced Koopman generator is identified directly from the lifted trajectory data.

\subsection{Controlled LWR Dynamics and Koopman Predictor}
To incorporate controlled vehicle entry into the traffic stream, we augment the conservation law with a spatially distributed source \cite{agarwal2015feedback},
\begin{equation}
    r_t
    +
    v_f\partial_x\!\left[r(1-r)\right]
    =
    \frac{u(t)}{\rho_j}b(x),
    \label{eq:controlled_lwr}
\end{equation}
where $u(t)$ denotes the total vehicle inflow and $b(x)$ specifies its spatial distribution along the road. We normalize the spatial profile such that $\int b(x)\,dx=1$, so that $u(t)$ represents the total vehicle inflow. A localized vehicle-entry point can be represented by choosing $b(x)$ to be concentrated over a small spatial region. In particular, a narrow Gaussian centered at a merge location provides a convenient representation of an on-ramp, in which case $u(t)$ corresponds to the admitted ramp flow. Similar Gaussian source profiles have been used to represent on-ramp influx in continuum traffic-flow models
\cite{LeeLeeKim1999}.

Applying the same polynomial observable $r^m$ to \eqref{eq:controlled_lwr}, and using
$\partial_t r^m=m r^{m-1}r_t$ together with
$\partial_x r^m=m r^{m-1}r_x$ and
$\partial_x r^{m+1}=(m+1)r^m r_x$, gives
\begin{equation}
    \partial_t r^m
    =
    -v_f\partial_x r^m
    +
    \frac{2m}{m+1}v_f\partial_x r^{m+1}
    +
    \frac{m}{\rho_j}b(x)u(t)r^{m-1}.
    \label{eq:controlled_hierarchy}
\end{equation}
The autonomous traffic dynamics therefore couple $r^m$ to the next-degree
observable $r^{m+1}$, whereas the ramp input acts through the lower-degree
observable $r^{m-1}$. Since $b(x)$ is prescribed, it acts as a fixed spatial multiplication operator and does not increase the polynomial degree of the control term. The constant observable $\psi_0=1$ is therefore included in the controlled
dictionary,
\(
    \widetilde{\boldsymbol{\Psi}}_p
    =
    \begin{bmatrix}
        1 & r & r^2 & \cdots & r^p
    \end{bmatrix}^{\top}.
\)
This ensures that the input term in the $m=1$ equation is represented within the retained observables. More generally, the control-dependent term in the $m$th equation involves only $r^{m-1}$, so it introduces no additional polynomial closure beyond the unresolved $r^{p+1}$ contribution identified in the autonomous dynamics.

After spatial discretization over $N_x$ cells, the polynomial observables are stacked into the lifted state
\begin{equation}
    \mathbf z
    =
    \begin{bmatrix}
        1 &
        \mathbf r^{\top} &
        (\mathbf r^{\circ 2})^{\top} &
        \cdots &
        (\mathbf r^{\circ p})^{\top}
    \end{bmatrix}^{\top}
    \in\mathbb{R}^{1+pN_x},
    \label{eq:lifted_state}
\end{equation}
where the elementwise-power notation was defined in the preceding subsection. The structure of \eqref{eq:controlled_hierarchy} motivates a bilinear dependence on the ramp input in the lifted coordinates. The reduced bilinear generator is identified directly from the lifted
traffic-density snapshots obtained from the spatio-temporal evolution of the LWR model.

For each sampled lifted state $\mathbf z_k$, a one-step successor
$\mathbf z_k^{+}$ separated by $\Delta t$ is used to approximate the
continuous-time derivative as
\[
    \dot{\mathbf z}_k
    \approx
    \frac{\mathbf z_k^{+}-\mathbf z_k}{\Delta t}.
\]
To reduce the dimension of the lifted model, the snapshot data are projected onto an orthonormal basis $U_r$. The constant observable is retained exactly in this basis, while the remaining directions are obtained from the dominant POD modes of the lifted snapshots. The reduced coordinates are
\[
    \mathbf a_k=U_r^{\top}\mathbf z_k,
    \qquad
    \dot{\mathbf a}_k=U_r^{\top}\dot{\mathbf z}_k .
\]

Normalizing the ramp input as $\mu_k=u_k/u_{\mathrm{scale}}$, we model the reduced dynamics as
\begin{equation}
    \dot{\mathbf a}
    =
    A_r\mathbf a
    +
    \mu N_r\mathbf a ,
    \label{eq:reduced_bilinear}
\end{equation}
where $A_r$ represents the autonomous dynamics and $N_r$ captures the
input-dependent contribution. For each training sample, define
\[
    \boldsymbol{\phi}_k
    =
    \begin{bmatrix}
        \mathbf a_k \\
        \mu_k\mathbf a_k
    \end{bmatrix},
\]
and collect \(
    \Phi
    =
    \begin{bmatrix}
        \boldsymbol{\phi}_1 & \cdots & \boldsymbol{\phi}_M
    \end{bmatrix},
    \qquad
    \dot{\mathcal A}
    =
    \begin{bmatrix}
        \dot{\mathbf a}_1 & \cdots & \dot{\mathbf a}_M
    \end{bmatrix}.
\)
The reduced generator matrices are obtained from the regularized regression
\begin{equation}
    \begin{bmatrix}
        A_r & N_r
    \end{bmatrix}
    =
    \dot{\mathcal A}\Phi^{\top}
    \left(
        \Phi\Phi^{\top}
        +
        \lambda I
    \right)^{-1},
    \label{eq:gedmd_regression}
\end{equation}
where $\lambda$ is the Tikhonov regularization parameter.

For a ramp command held constant over a control interval $T_c$,
\eqref{eq:reduced_bilinear} gives the finite-time predictor
\begin{equation}
    \mathbf a_{k+1}
    =
    \exp\!\left[
        \left(A_r+\mu_kN_r\right)T_c
    \right]\mathbf a_k.
    \label{eq:finite_time_predictor}
\end{equation}
The matrix exponential provides a finite-horizon predictor that can be used to evaluate how the traffic state evolves under different input choices. Later in this paper, this predictor is used within the MPC framework for the ramp-metering case study.

\subsection{Koopman-Based Receding-Horizon Control}\label{subsec:controller}

In this subsection we present how the Koopman model can be embedded within a receding-horizon control framework. At each control update, a sequence of candidate inputs is propagated over a prediction horizon of $H$ intervals using \eqref{eq:finite_time_predictor}. The corresponding traffic-density fields are reconstructed from the reduced lifted states and used to evaluate the effect of each candidate input sequence. Let $\mathbf{u}_k=[u_k,u_{k+1},\ldots,u_{k+H-1}]^{\top}$ denote a candidate control sequence. A general finite-horizon optimization
can then be written as
\vspace{-2mm}
\begin{equation}
    \min_{\mathbf{u}_k}
    \;
    J_k
    =
    \sum_{j=1}^{H}
    \ell\!\left(
        \widehat{\mathbf r}_{k+j},
        u_{k+j-1},
        u_{k+j-2}
    \right),
    \label{eq:generic_mpc}
\end{equation}
\vspace{-1mm}
subject to the Koopman predictor and any constraints imposed by the
particular traffic-control problem. Here, $\widehat{\mathbf r}_{k+j}$ denotes the predicted normalized traffic-density field and $\ell(\cdot)$ is a stage cost that defines the desired control objective.

Depending on the application, the stage cost may penalize quantities such as high traffic density, operation above a critical density, restricted vehicle entry, or rapid changes in the control input. Similarly, the admissible input set and other constraints can be selected according to the physical requirements of the traffic-control problem. The formulation therefore allows the same Koopman predictor to be used with different control objectives without changing the underlying reduced-order traffic model.

In the following section, this general framework is demonstrated for a
freeway ramp-metering problem, where the objective function and control
constraints are specified for that application.

\section{Ramp-Metering Case Study}
\label{sec:case_study}

We evaluate the proposed Koopman predictor in a freeway
ramp-metering case study. The Koopman model is first tested on an
unseen traffic trajectory and is then used within a receding-horizon
controller. Its performance is compared with predictors based on a fixed
linearization of the LWR model and direct nonlinear finite-volume
propagation. All controllers use the same nonlinear freeway model, prediction horizon, admissible inputs, and objective function; only the internal prediction model is changed.

\subsection{Experiment Design}

The case study considers a $5~\mathrm{km}$ freeway governed by the LWR model with the Greenshields flow--density relation. The free-flow speed and jam density are $v_f=100~\mathrm{km/h}$ and
$\rho_j=150~\mathrm{veh/km}$, respectively. The normalized upstream and
downstream boundary densities are fixed at $r_{\mathrm{in}}=0.30$ and
$r_{\mathrm{out}}=0.08$. These parameters give a Greenshields capacity of
$3750~\mathrm{veh/h}$ and an upstream equilibrium flow of
$3150~\mathrm{veh/h}$, leaving approximately $600~\mathrm{veh/h}$ of
available capacity before additional vehicle inflow is introduced. The
on-ramp is located at $x=2~\mathrm{km}$ and is represented by the localized
Gaussian source described in Section~\ref{sec:koopman_lwr}. To assess the role of the prediction model, we compare three controllers that
differ only in their internal predictor: (i) the proposed reduced Koopman model, (ii) a fixed linearized LWR model, and (iii) the nonlinear finite-volume LWR model.

\begin{figure}[!htbp]
    \centering
    \includegraphics[width=0.48\textwidth]{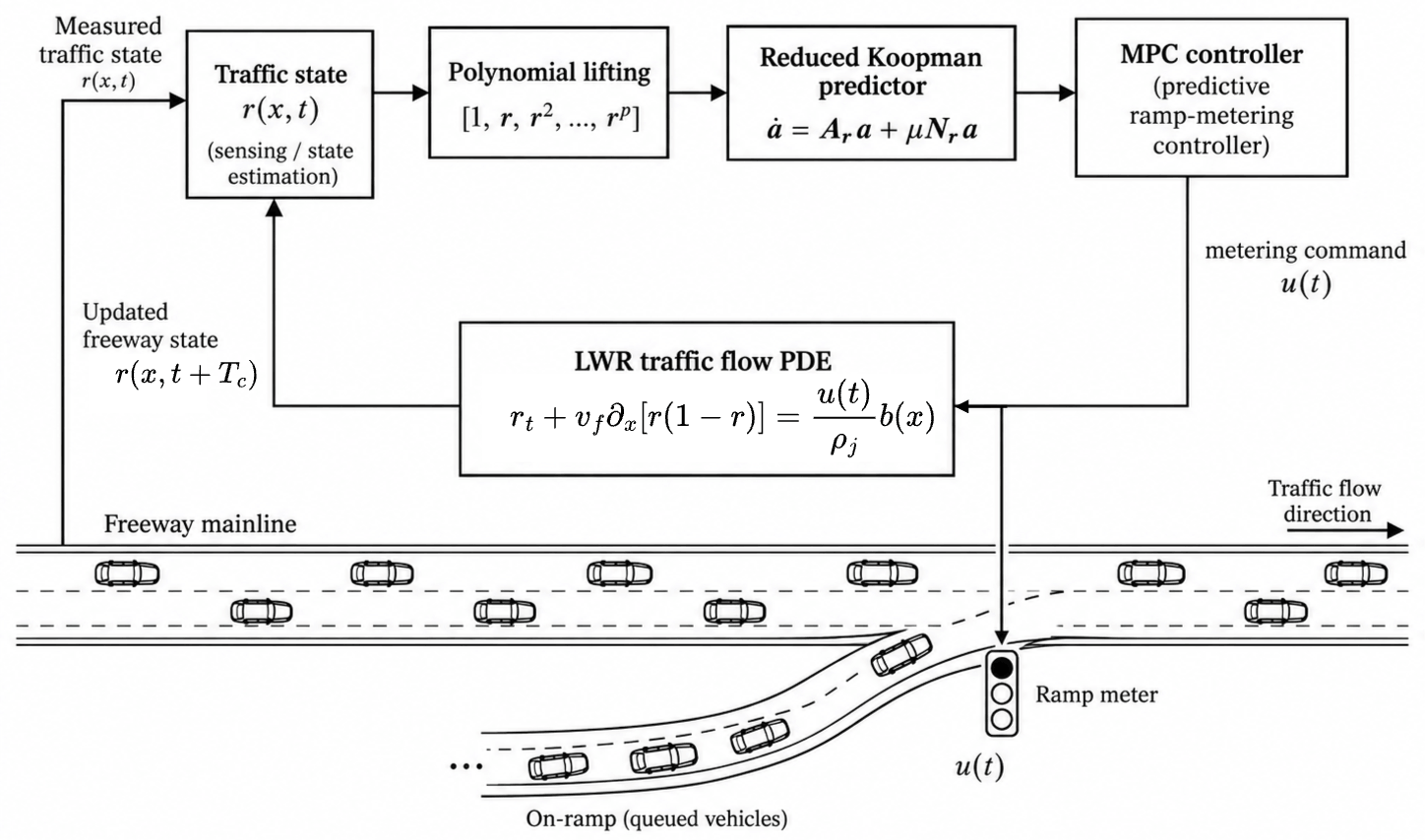}
    \caption{Conceptual workflow of the Koopman-based predictive control framework. The measured traffic-density field is lifted to polynomial observables and propagated using the reduced Koopman predictor. The predicted traffic evolution is used to determine the control input and applied to the nonlinear LWR freeway model}
    \label{fig:conceptual_framework}
    \vspace{-4mm}
\end{figure} 

Figure~\ref{fig:conceptual_framework} illustrates the overall workflow of the case study. At each control update, the current freeway density field
$r(x,t)$ is assumed to be available from traffic sensing or state-estimation methods. The measured density is mapped to the polynomial observable space and then projected onto the reduced Koopman coordinates. The Koopman model is used to predict how the traffic state evolves under different candidate control inputs. These predictions are supplied to the controller, which selects an input
based on the expected future traffic conditions. For the present case study, the input corresponds to the metered vehicle flow entering from the on-ramp. The selected control input is applied to the nonlinear LWR traffic model, which represents the actual freeway dynamics and is referred to as the \emph{plant} in the control framework. The resulting traffic-density field becomes the updated state for the next control step, closing the feedback loop. The same nonlinear freeway model is used as the plant in all controlled cases, while only the prediction model inside the controller is changed. This allows the proposed Koopman, fixed linearized LWR, and nonlinear finite-volume LWR predictors to be compared under identical traffic and control settings. The numerical simulation and implementation details are presented in the next subsection.

\subsection{Numerical Simulation and Controller Setup}
We will compare the controllers under a time-varying ramp-demand scenario. The simulation spans one hour. The ramp demand is specified as
$400$, $800$, $1400$, $600$, $1200$, $400$, $1600$, $1000$, and
$400~\mathrm{veh/h}$ over the consecutive intervals
$0$--$5$, $5$--$10$, $10$--$18$, $18$--$23$, $23$--$33$, $33$--$38$,
$38$--$50$, $50$--$55$, and $55$--$60~\mathrm{min}$, respectively. Four cases are considered: (i) an uncontrolled case in which the full ramp
demand is admitted, (ii) Koopman-MPC, (iii) linearized LWR-MPC, and (iv) nonlinear LWR-MPC. The same nonlinear finite-volume LWR model governs the actual freeway traffic in all cases. The three controlled cases differ only in the model used to predict future traffic conditions. Koopman-MPC uses the proposed reduced Koopman predictor, linearized LWR-MPC uses a fixed affine model linearized about $r=0.43$, and nonlinear LWR-MPC directly propagates the nonlinear finite-volume model over the prediction horizon. 

The LWR model is discretized using $N_x=80$ finite-volume cells
($\Delta x=0.0625~\mathrm{km}$) and propagated using a Rusanov flux with
third-order strong-stability-preserving Runge--Kutta (SSPRK3) integration and a $1~\mathrm{s}$ time step. The Koopman model is trained from 25 independent $10~\mathrm{min}$ trajectories with inputs held for $30~\mathrm{s}$; snapshots are retained every $2~\mathrm{s}$, yielding 7500 training samples. The Koopman predictor uses polynomial observables up to degree $p=5$, resulting in $401$ lifted coordinates, which are reduced to $40$ coordinates for prediction and control. The controller is updated every $T_c=30~\mathrm{s}$ and uses $H=10$ prediction intervals, corresponding to a $5~\mathrm{min}$ horizon. To reduce the number of candidate input sequences, two-block move blocking is used. One input is held over the first five prediction intervals and a second input over the remaining five. All three controllers use the same move-blocking structure and the same admissible input grid $\mathcal U=\{0,200,\ldots,1600\}\,\mathrm{veh/h}$.

The generic stage cost introduced in Section~\ref{subsec:controller} is specialized to penalize four quantities: (i) density above the desired operating level ($w_h=30$), (ii) density above the critical value ($w_c=350$), (iii) withheld ramp demand ($w_u=0.35$), and (iv) changes in the metering input ($w_{\Delta u}=0.05$). All weights $(w_h,w_c,w_u,w_{\Delta u})$ are kept identical for the three
controllers and selected empirically to place the strongest emphasis on avoiding operation above the critical density while also accounting for ramp admission and smooth metering actions. Weighted traffic-performance objectives with penalties on control variation are commonly used in freeway MPC \cite{Bellemans2006}. The first two terms are evaluated as mean-squared density exceedances over the control region, while the latter two are normalized by the maximum admissible ramp flow. For this experiment, the desired operating level is $r_{\mathrm{tar}}=0.43$ and the critical density is $r_c=0.50$, and the density penalties are evaluated over the region extending from $0.35~\mathrm{km}$ upstream to $1.20~\mathrm{km}$ downstream of the ramp.

At each control update, only inputs that do not exceed the forecast ramp demand are considered. The control sequence with the lowest predicted cost is selected, and only its first input is applied before the traffic state is updated and the optimization is repeated. 

\subsection{Closed-Loop Results}
Table~\ref{tab:control_results} provides a quantitative comparison of the
closed-loop performance of the three MPC formulations and the uncontrolled case. Without ramp control, the maximum normalized density reaches $0.83$. Koopman-MPC reaches a maximum density of $0.472$, compared with $0.468$ for nonlinear LWR-MPC, corresponding to a difference of only $0.83\%$.

\vspace{-2mm}
\begin{table}[!htbp]
\centering
\caption{Closed-loop performance over the ramp-metering scenario.}
\label{tab:control_results}
\footnotesize
\setlength{\tabcolsep}{2.5pt}
\renewcommand{\arraystretch}{1.08}

\begin{tabularx}{\columnwidth}{
@{}
>{\raggedright\arraybackslash}X
cccc
@{}
}
\hline
Metric
& Uncon-.
& Koopman
& Linearized
& Nonlinear \\
& trolled
& MPC
& LWR-MPC
& LWR-MPC \\
\hline

Maximum density
& 0.83 & 0.472 & 0.42 & 0.468 \\

Admitted ramp vehicles
& 1007 & 513 & 392 & 510 \\

Mean computation [ms]
& -- & 2.732 & 0.589 & 32.492 \\

Total computation [s]
& -- & 0.328 & 0.071 & 3.899 \\

\hline
\end{tabularx}
\end{table}
\vspace{-2mm}

The similarity between Koopman-MPC and nonlinear LWR-MPC is also reflected in the admitted ramp traffic. Koopman-MPC admits $513$ vehicles during the one-hour experiment, compared with $510$ vehicles for nonlinear LWR-MPC. The difference is less than $1\%$. The fixed linearized LWR controller also prevents congestion, but admits only $392$ ramp vehicles. Its lower maximum density of $0.42$ therefore results from a substantially more restrictive metering policy. This behavior is consistent with the larger prediction errors observed for the fixed linearized model away from its operating point. The three prediction models exhibit a clear trade-off between computational cost and model fidelity. The fixed linearized LWR predictor is the least expensive, requiring only $0.589~\mathrm{ms}$ per control decision, but its limited prediction accuracy produces a noticeably different control policy. At the other extreme, nonlinear LWR-MPC provides the highest-fidelity prediction by repeatedly integrating the finite-volume model, but requires $32.492~\mathrm{ms}$ per control decision. Koopman-MPC provides an intermediate computational cost of $2.732 ~\mathrm{ms}$ while closely reproducing both the density response and ramp admission obtained with the nonlinear controller. Its mean online computation is reduced by a factor of approximately $12$ relative to nonlinear LWR-MPC. For the present freeway example, all three controllers remain well within the $30~\mathrm{s}$ control-update interval. The advantage of the Koopman model is that it offers a favorable compromise between the computational efficiency of a linear predictor and the predictive fidelity of the nonlinear LWR model.

\begin{figure}[!htbp]
    \centering
    \includegraphics[width=0.48\textwidth]{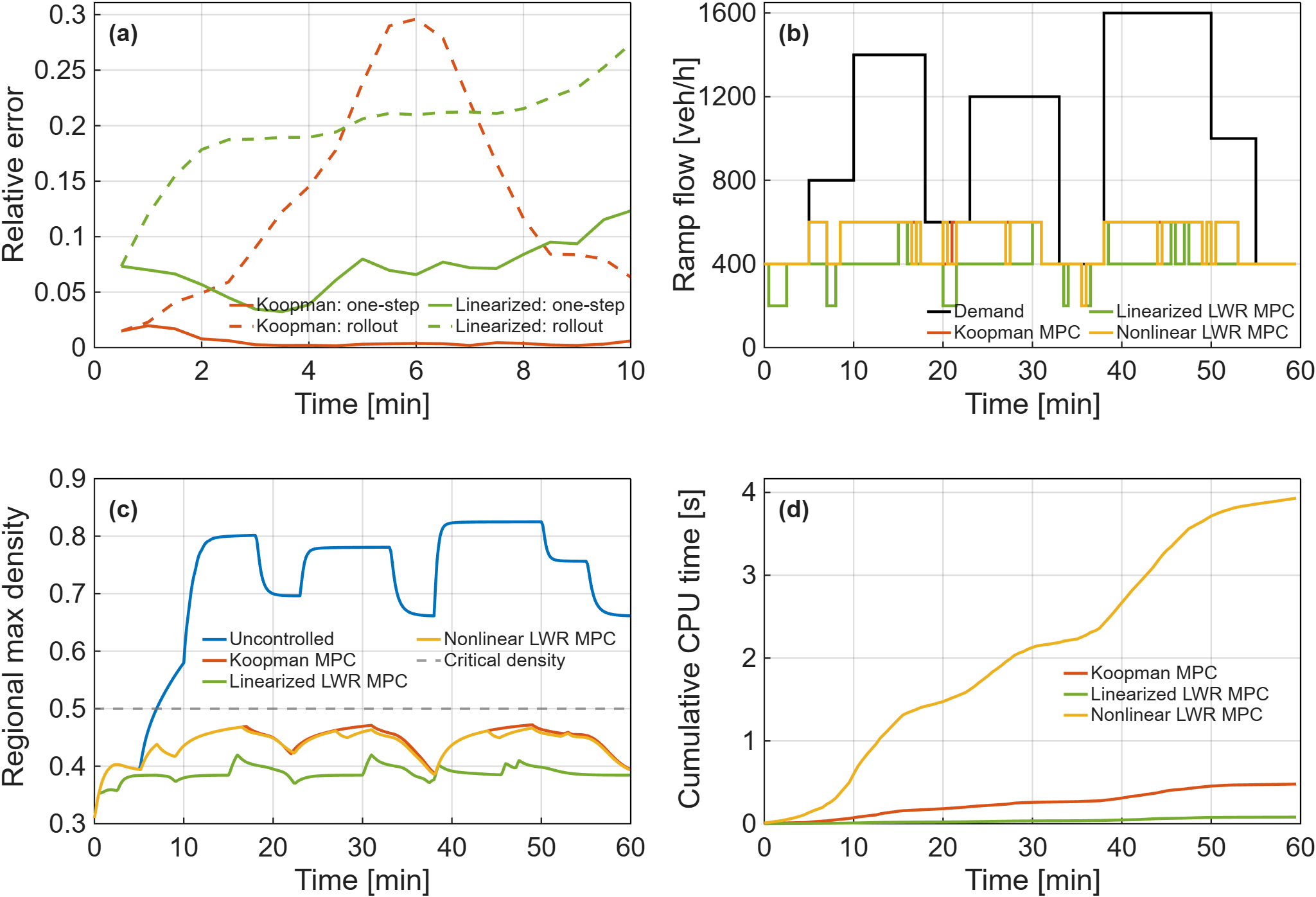}
    \caption{Ramp-metering comparison. (a) One-step and recursive rollout errors of the reduced Koopman and fixed linearized LWR predictors relative to the nonlinear finite-volume LWR model. (b) Ramp demand and metering inputs selected by the three MPC controllers. (c) Maximum normalized density in the ramp region for the uncontrolled and controlled cases; the dashed line marks the critical density $r_c=0.5$. (d) Cumulative online computation time of the three controllers.}
    \label{fig:case_results}
    %\vspace{-8mm}
\end{figure}

Figure~\ref{fig:case_results} compares the prediction accuracy, control actions, traffic response, and computational cost of the three MPC formulations. In Fig.~\ref{fig:case_results}(a), the prediction error is
computed as the relative Euclidean error
$\|\hat{\mathbf r}-\mathbf r\|_2/\|\mathbf r\|_2$, where $\mathbf r$ is the nonlinear finite-volume LWR solution. The one-step error is evaluated by reinitializing each predictor from the current true state, whereas the recursive rollout error is obtained by repeatedly propagating the predicted
state. The proposed Koopman model remains substantially closer to the nonlinear LWR reference than the fixed linearized model, particularly over recursive prediction. The uncontrolled case develops a sustained high-density region near the ramp, while all three MPC controllers keep the regional maximum density below the critical value. Koopman-MPC closely follows the nonlinear LWR-MPC response and admits a similar amount of ramp traffic. In contrast, the linearized LWR-MPC achieves the lowest computational cost but produces larger prediction errors and a more restrictive metering policy, resulting in fewer admitted ramp vehicles. Thus, the linearized model favors computational speed at the expense of prediction fidelity and ramp admission, whereas the nonlinear LWR model provides the most accurate prediction at the highest online cost. The Koopman predictor provides a useful compromise between these two extremes by retaining the nonlinear traffic behavior needed for control
while substantially reducing repeated online model evaluation cost.

\section{Conclusion}
\label{sec:conclusion}

This study showed that the analytical structure of the LWR model can be used to guide the selection of Koopman observables instead of relying entirely on heuristic or learned lifting functions. For the Greenshields relation, the resulting quadratic flux leads naturally to polynomial density observables, and the controlled formulation leads to a reduced bilinear model. The ramp-metering study highlights the main trade-off among the three prediction models considered. The fixed linearized LWR model is the least expensive computationally, but its prediction errors are substantially larger and it produces a more restrictive metering policy. Direct nonlinear LWR prediction provides the highest fidelity but at the largest online cost. The Koopman predictor lies between these two extremes: it closely reproduces the nonlinear MPC response and ramp admission while reducing the repeated online computation by approximately one order of magnitude.

As a proof of concept for the proposed polynomial Koopman observables, the present study focuses on a freeway scenario. The case study uses the Greenshields flow--density relation, a single on-ramp, and synthetic traffic data. The case study demonstrates the potential of the proposed framework in a controlled setting and evaluates the effectiveness of the Koopman-based predictor. Future work will extend the framework to more complex freeway configurations, other flow--density relations, explicit ramp-queue dynamics, and real traffic data. We also plan to investigate additional traffic prediction and control problems. More broadly, the same physics-guided Koopman idea will be explored for other systems governed by nonlinear conservation laws, where the structure of the flux may provide useful guidance for selecting observables.

\bibliographystyle{IEEEtran}
\bibliography{ACC_ref}

\end{document}